\documentclass{aa}
\usepackage{graphicx}
\usepackage{rotating}
\begin{document}
\title{{\it BeppoSAX} observations of the quasar Markarian~205}
\author{P. Favre\inst{1,2}
\and 
P.-O. Petrucci\inst{3}
\and
V. Beckmann\inst{4,5}
\and
T. J.-L. Courvoisier\inst{1,2}
}
\offprints{Pascal.Favre@obs.unige.ch}
\institute{INTEGRAL Science Data Centre, 16 Ch. d'Ecogia, 1290 Versoix, Switzerland
\and
Observatoire de Gen\`{e}ve, 51 Ch. des Maillettes, 1290 Sauverny, Switzerland
\and
Laboratoire d'Astrophysique de Grenoble, BP 43, 38041
Grenoble Cedex 9, France 
\and
NASA Goddard Space Flight Center, Code 661, Greenbelt, MD
20771, USA
\and
Joint Center for Astrophysics, Department of Physics, University of Maryland, Baltimore County, Baltimore, MD 21250, USA}
\date{Received ; accepted }
\abstract{We present the first {\it BeppoSAX} observation (0.1 to 220
  keV) of the quasar \object{Mrk~205}. We have searched for the
  unusual Fe line profile claimed in the {\it XMM-Newton} spectrum
  which has been widely discussed in recent literature. We find no
  evidence for a broad, ionized Fe line component in our data. We 
detect for the first time a Compton hump in this object. Besides, when
  this component is included in the fit, the line strength diminishes,
  in agreement with a recent re-analysis of the
  {\it XMM-Newton} data, but with better constraints on the reflection
  component thanks to the PDS instrument (15-220 keV). We interpret this
  fact as another indication for illumination of a distant and cold
  material rather than reprocessing in the highly ionized inner parts of
  an accretion disk. \\
We cannot constrain the presence of a high energy cutoff but we confirm the existence of a variable soft excess (one year timescale).  
}
\titlerunning{{\it BeppoSAX} observation of \object{Mrk~205}} \maketitle
\section{Introduction}
Discovered by Weedman (\cite{weedman}), the quasar \object{Mrk~205} is
seen through the southern spiral arm of the nearby spiral, barred galaxy
\object{NGC~4319} at a projected distance of roughly 40\arcsec\, from its
center. From its radio luminosity at 6 cm, \object{Mrk~205} is classified
as a radio-quiet quasar (Rush et al. \cite{rush}), but it is still not
clear whether it belongs to the radio-quiet quasar class (e.g. Bahcall et
al. \cite{bahcall}) or to the Seyfert~1 class (e.g. Veron-Cetty \& Veron
\cite{veron}). \object{Mrk~205} has a redshift $z=0.07085$ (Huchra et al.
\cite{huchra}) while \object{NGC~4319} has $z=0.00468$ (Bowen et
al. \cite{bowen1}). \\
\\
In the standard paradigm of (radio-quiet) Active Galactic Nuclei (AGN),
X-ray photons result from Compton upscattering of optical-UV photons in a
hot thermal corona above the accretion disk surrounding a supermassive
black hole. A large fraction of the seed photons are thought to be
produced by thermal emission of the accretion disk itself, but
\object{Mrk~205} was found by McDowell et al. (\cite{mcdowell}) to have a
weak blue bump.  One of the possible scenarios which could explain this
spectral morphology is that the bump may be highly variable and thus
observed in a particularly weak state. In the thermal comptonisation
  framework, blue bump variability would also induce hard X-ray
variations.\\
A soft X-ray excess component may be interpreted as the hard tail of 
the blue bump (Walter \& Fink \cite{walterfink}; Brunner et al. 
\cite{brunner}). So far, evidence for a soft excess in \object{Mrk~205} 
was only reported in two observations; 
with {\it EINSTEIN} IPC (Wilkes \& Elvis
\cite{wilkeselvis}) and {\it XMM-Newton} (Reeves et
al. \cite{reeves2}).\\
{\it BeppoSAX}, with its unique broad-band capabilities allows to extend
the observation of \object{Mrk~205} for the first time above 20 keV to
study the properties of the hard X-ray emission as well as of a potential
reflection component. Indeed, some of the X-ray photons can be
reprocessed in the surroundings, in particular in the disk itself giving
rise to neutral (or low-ionization) Fe~K$\alpha$ line.  While it is a
common feature of Seyfert~1 galaxies (Nandra \& Pounds \cite{nandra}),
only a few quasars present a clear detection of such a component
(Williams et al.  \cite{williams}; Nandra et al. \cite{nandra2}). 
Furthermore, the centroid of the line
was often measured at energies close to 6.7 keV suggesting that its
origin may be in an ionized layer of the accretion disk (Reeves \& Turner
\cite{reeves1}), the neutral Fe~K$\alpha$ being emitted in distant and
cold material lying outside the broad line region. Reeves et al.
(\cite{reeves2}), who analysed the {\it XMM-Newton} data of
\object{Mrk~205}, claimed the existence of a broad ionized component in
addition to the neutral one in this object. However, Page et al. 
(\cite{page}) reprocessed the data including a reflection hump
component in the fit and found little evidence for a 
broad, ionised component in the data. \\
In this paper, we present the analysis of three {\it BeppoSAX}
observations of \object{Mrk~205} for a total of 200 ks of exposure time. In
Section 2, we summarize the knowledge we had on \object{Mrk~205} prior to
our {\it BeppoSAX} observation. In Section 3, we describe the data and
their analysis while in Section 4 we compare our results and discuss them
in the framework of the unified model.
\section{Previous X-ray observations of \object{Mrk~205}}
\begin{figure}
\includegraphics[width=9cm]{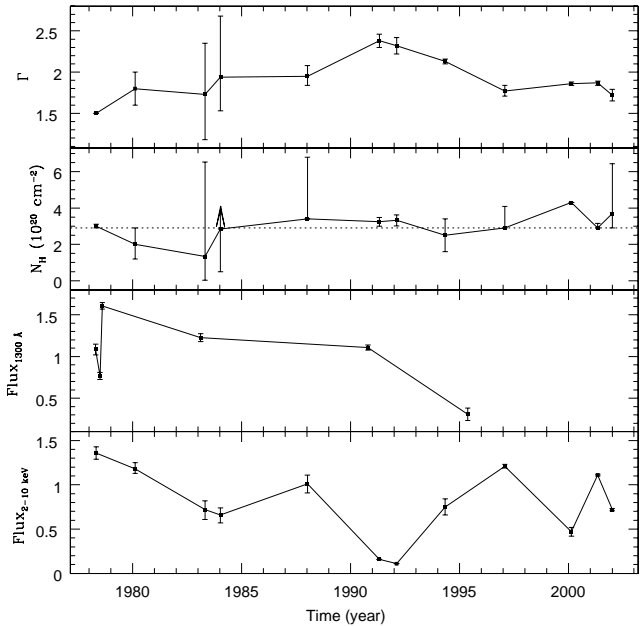} 
\caption{ Summary of the X-ray observations of \object{Mrk~205};
  evolution with time of the photon index $\Gamma$, the Hydrogen
  column density in the line of sight $N_H$ (the 
  dotted line denotes the Galactic value) and the $2-10$ keV flux; 
units 10$^{-11}$ erg cm$^{-2}$ s$^{-1}$. Data from Table~\ref{Tab1}.
  We have included as well an {\it IUE} light curve at 1300 \AA\,; 
units 10$^{-14}$ erg cm$^{-2}$ s$^{-1}$ \AA$^{-1}$}
\label{Fig1}
\end{figure}
\begin{table*}[tb]
\caption{ Summary of the X-ray observations of \object{Mrk~205}; spectral parameters.
The $2-10$ keV flux was extrapolated from the best fit parameters
}\label{Tab1}
\begin{flushleft}
\addtolength{\tabcolsep}{-2pt}
\begin{tabular}{@{}lcccc@{}}
\hline
\hline
\rule[-0.5em]{0pt}{1.6em}
Mission, Reference & Date & $\Gamma$ & $N_H$ & Flux$_{2-10 {\rm keV}}$\\
\hline  
\rule[-0.5em]{0pt}{1.6em}      
&&& 10$^{20}$ cm$^{-2}$ & 10$^{-11}$erg cm$^{-2}$ s$^{-1}$  \\
\hline
{\it EINSTEIN (HRI)}, [a]&22-11-1978&1.50$^{f} $&3.00$^{+0.10}
_{-0.10} $&1.36$^{+0.07} _{-0.07}$\\
{\it EINSTEIN (IPC)}$^{\dagger}$, [b]&20-04-1980&1.80$^{+0.20} _{-0.20}$&2.00$^{+0.90}
_{-0.80}$&1.18$^{+0.07} _{-0.05}$\\
{\it EXOSAT}$^{\star}$, [c]&10-11-1983&1.73$^{+0.62} _{-0.55}$&1.33$^{+5.20}
_{-1.30}$&0.72$^{+0.10} _{-0.11}$\\
{\it EXOSAT}$^{\star}$, [c]&27-01-1984&1.94$^{+0.74} _{-0.41}$&2.84$^{+13.70}
_{-2.35}$&0.66$^{+0.08} _{-0.09}$\\
{\it GINGA}$^{\ddagger}$, [d]&16-01-1988&1.95$^{+0.13}
_{-0.11}$&3.40$^{+3.40} _{-0}$&1.01$^{+0.10} _{-0.10}$\\
{\it ROSAT (PSPC)}, [e]&10-11-1991&2.38$^{+0.08} _{-0.08}$&3.24$^{+0.24}
_{-0.24}$&0.16$^{+0.01} _{-0.01}$\\
{\it ROSAT (PSPC)}, [e]&15-04-1992&2.32$^{+0.10} _{-0.10}$&3.32$^{+0.30}
_{-0.30}$&0.11$^{+0.01} _{-0.01}$\\
{\it ASCA}$^{\ddagger}$, [f]&03-12-1994&2.13$^{+0.03} _{-0.03}$&2.50$^{+0.90}
_{-0.90}$&0.75$^{+0.09} _{-0.09}$\\
{\it BeppoSAX}, this work&12-03-1997&1.77$^{+0.07} _{-0.06}$&2.90$^{+1.18}
_{-0}$&1.21$^{+0.02} _{-0.02}$\\
{\it XMM-Newton}$^{\dagger}$$^{\ddagger}$,
[g]&06-05-2000&1.86$^{+0.02} _{-0.02}$&4.30 &0.47$^{+0.05} _{-0.05}$\\
{\it BeppoSAX}$^{\dagger}$$^{\ddagger}$, this work&21-12-2001&
1.87$^{+0.02} _{-0.03}$&2.90$^{+0.25}
_{-0}$&1.11$^{+0.01} _{-0.01}$\\
{\it BeppoSAX}, this work&05-01-2002&1.72$^{+0.07}
_{-0.07}$&3.69$^{+2.75} _{-0.79}$&0.72$^{+0.01} _{-0.01}$\\
\hline
\end{tabular}
\end{flushleft}
Note: Galactic $N_H=2.90\times10^{20}$ cm$^{-2}$ (Dickey \& Lockman \cite{dickey}) \\
$^{\dagger}$ evidence for a soft excess component reported\\
$^{\star}$ broken power law model preferred\\
$^{\ddagger}$ evidence for an Fe K$\alpha$ line reported\\
$^f$ fixed parameter
\end{table*}

\object{Mrk~205} was observed by {\it EINSTEIN} HRI (Tananbaum et al. 
\cite{tananbaum}: [a]; Zamorani et al. \cite{zamorani}), {\it
  EINSTEIN} IPC (Wilkes \& Elvis \cite{wilkeselvis}: [b]),
{\it EINSTEIN} A-2 (Della Ceca et al. \cite{dellaceca}), {\it EXOSAT} 
(Singh et al. \cite{singh}: [c]), {\it GINGA} 
(Williams et al. \cite{williams}: [d]), {\it ASCA} (Reeves \& Turner 
\cite{reeves1}: [f]), {\it ROSAT} PSPC (Fiore et al. \cite{fiore4}:[e]; 
Fiore \cite{fiore0}; Rush et al. \cite{rush2}; Ciliegi \& Maccacaro 
\cite{ciliegi}; Fiore et al. \cite{fiore3}), {\it ROSAT} HRI 
(Arp \cite{arp96}), {\it XMM-Newton} (Reeves et al. \cite{reeves2}:
[g]; Page et al. \cite{page}) and {\it BeppoSAX} (this work). \\
\\
\object{Mrk~205} was neither observed by {\it RXTE} nor {\it Chandra} to date. \\
\\
We summarize the spectral parameters found in Fig.~\ref{Fig1} and
Table~\ref{Tab1}. The spectral variability was important over the 24
years spanned by the data (Fig.~\ref{Fig1}, upper panel). The photon
indexes were measured mostly over the range $2-10$ keV and thus were not 
affected by a soft excess component. However, the 
two {\it ROSAT} points (1991, 1992) are extrapolated values from the
$0.11-2.45$ keV, i.e. a spectral range very sensitive to the presence
of a soft excess. They are certainly overestimated. In addition,
several papers
discuss the {\it ROSAT} PSPC calibration problems which results in a measure 
of the AGN X-ray spectral slopes being significantly steeper than those from
other missions (Beckmann et al. \cite{beckmann} and references therein).\\
\noindent
We will discuss further in Section 4 the possible origin of the
  observed spectral variability in this object.\\
Little variations of the Hydrogen column density were observed during the
same period (Fig.~\ref{Fig1}, second panel from the top).  An excess of
absorption may be present in the {\it XMM-Newton } 2000 observation, its
reality will be discussed in Section 3.7.
\\
\noindent  
The $2-10$ keV flux seems to have decreased during the first $\sim$ 18
years of observation, a trend possibly correlated with a similar trend
seen in the 1300 \AA\, {\it IUE} light curve (Fig.~\ref{Fig1}, third
panel from the top). Here again, the two {\it ROSAT} estimates of 
the $2-10$ keV flux may be underestimated due to the overestimation of
 the $2-10$ keV photon index.
\noindent   
In the remainder of this section, we will focus on the {\it
  XMM-Newton} observation which took place one and a half year before 
our {\it BeppoSAX} pointings. 

\subsection{{\it XMM-Newton} observation}
Reeves et al. (\cite{reeves2}) reported an observation done with
{\it XMM-Newton} on 6 May 2000. They measured a photon index
$\Gamma=1.86^{+0.02} _{-0.02}$ with an absorbing column in excess of the
Galactic one limited to $N_H$ $<$ $1.5\times 10^{20}$ cm$^{-2}$. Fitting
the MOS data only they found evidence for an ionized Fe line component at
6.7 keV along with the neutral Fe K$\alpha$ component. The same result
was obtained using only the PN data, and thus the simultaneous fit of the
data from both instruments gave tight constraints. They finally obtained
the best fit with a narrow line resulting from neutral Fe K$\alpha$
emission at $6.39\pm 0.02$ keV ($EW=56\pm 23$ eV) and a broad component
at $6.74\pm 0.12$ keV ($EW=135^{+65} _{-60}$ eV). They did not detect Fe
K edge, nor a reflection component but claimed that they could not rule
out its presence.\\
The observations were consistent with a relativistically-broadened
accretion disk line model (Fabian et al. \cite{fabian}) only if the 
disk ionization was high enough to produce
He-like Fe at 6.7 keV and the neutral component was
present.\\
A clear soft excess was also detected, but fitting it with a black body
gave a much too high temperature ($kT=120\pm 8$ eV) to be direct emission
from the accretion disk. An explanation is enhanced emission from the
upper layers of the ionized accretion disk. The fact that the Compton
reflection from ionized material model ({\sc pexriv}: Magdziarz \&
Zdziarski \cite{magdz}) gave good results (with $\xi=300\pm80$ erg cm
s$^{-1}$) was probably confirming this
(see also Nayakshin et al. \cite{nayaketal}).  No warm absorber was
found; the limits on the OVII
and OVIII edges were $\tau <0.12$ and $\tau <0.15$ respectively.\\
\begin{figure*}
\sidecaption
\includegraphics[width=17cm]{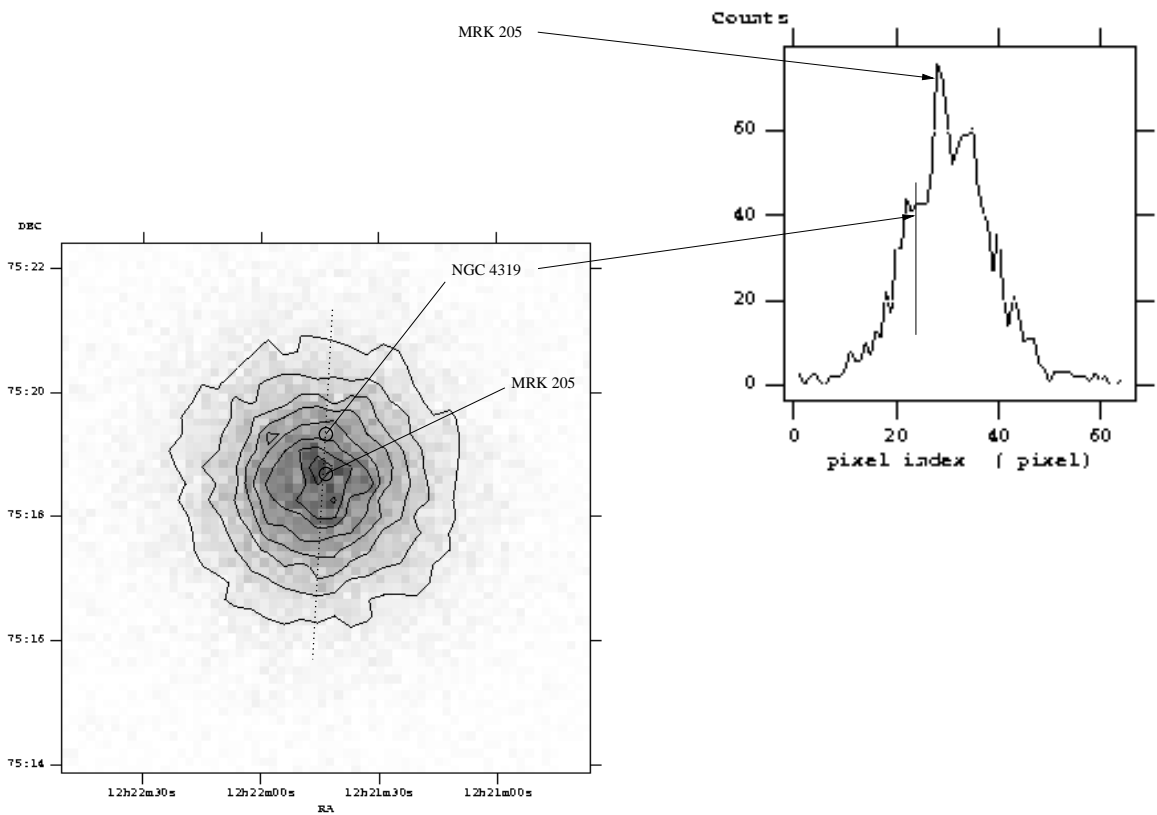} 
\caption{ Left panel: MECS intensity image of the 2001 observation with contours
  at (from the outside) 7.7, 15.5, 23.3, 31.1, 38.8, 46.6, 54.4, 62.2 counts.  
 Right panel: profile along the dotted line displayed on the left
 panel (1 MECS pixel is 7\arcsec). \object{NGC~4319} is located 6 pixels on the left of the
    central peak in a region which accounts for 54\% of the peak
 value}
\label{Fig2}
\end{figure*}
Page et al. (\cite{page}) reprocessed these data and found
that an acceptable fit was obtained with a power law plus neutral emission line
at 6.4 keV. They noticed that a better fit was obtained either with a
Compton reflection component or with an ionized emission line, with a
similar goodness of fit. The presence of ionized relativistic material
was thus not unambiguously
detected in this object, the illumination of a distant cold material
providing a simpler self-consistent explanation.
\section{{\it BeppoSAX} observation}
\object{Mrk~205} was observed by {\it BeppoSAX} (Boella et al.
\cite{boella1}) on 21 December 2001 (LECS:~47\,582~s, MECS:~165\,458~s,
PDS:~70\,549~s) and on 5 January 2002 (LECS:~3\,611~s, MECS:~32\,603~s,
PDS:~12\,606~s). In addition, public data from the archive (12 March
1997, LECS:~2\,525~s, MECS:~14\,727~s, PDS:~6\,646~s; part of the core
program and pointed on \object{NGC~4319}) are
included in the present study.\\
LECS (Low Energy Concentrator Spectrometer, Parmar et al. \cite{parmar})
and MECS (Medium Energy Concentrator Spectrometer, Boella et al.
\cite{boella2}) data were reduced following the standard procedures. For
PDS (Phoswich Detector System, Frontera et al. \cite{frontera}) data,
high level products from the Sax Data Center were used.  LECS and MECS
data have been screened according to Fiore et al. (\cite{fiore}) to
produce equalized event files using an extraction radius of 8\arcmin\,
for LECS and 4\arcmin\, for MECS. Data from the two MECS units have been
merged to increase the signal-to-noise ratio.
\\
For the longest exposure (2001), the average source flux in the $2-10$
keV band (MECS data only, power law model with Galactic absorption) was
$(1.11\pm 0.01)\times 10^{-11}$ erg cm$^{-2}$ s$^{-1}$ corresponding to a
luminosity $L=2.55\times 10^{44}$ erg s$^{-1}$ in the $2-10$ keV band
rest frame (here and in the following $H_0=50$ km s$^{-1}$ Mpc$^{-1}$). From that
luminosity, a black hole mass of about $2\times 10^{6}$
$M_{\odot}L_{Edd}/L$ could be inferred. These values are more typical of
a luminous Seyfert~1 galaxy than a quasar.
\noindent
The $2-10$ keV flux measured in 2001 was twice the value measured one 
and a half year before by {\it XMM-Newton} (Fig.~\ref{Fig1}).

\subsection{Image analysis: unprobable influence from \object{NGC~4319}
  and other sources} 
\begin{figure}
\includegraphics[height=9cm,angle=-90]{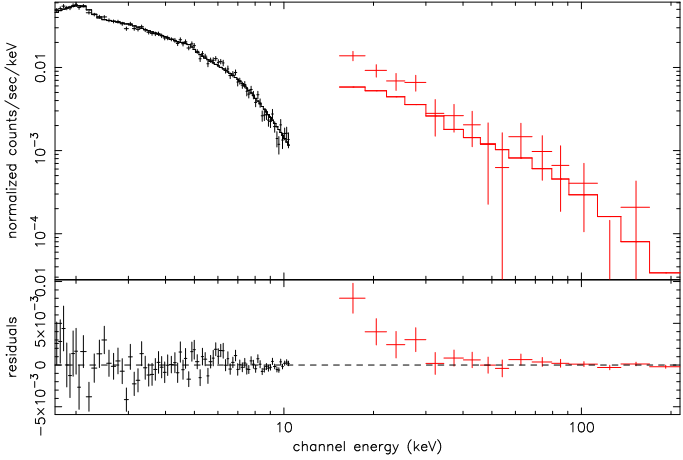} 
\caption{ Upper panel: fit of the 2001 MECS data with a power law and absorption frozen to
the Galactic value. PDS data are included without being fitted but
with an intercalibration constant frozen to 0.88. 
Lower panel: residuals with respect 
to the MECS best fit model}
\label{Fig2b}
\end{figure}

We analysed the MECS $1.3-10$ keV image (angular
resolution 42\arcsec) of the 2001 observation and compared it with a {\it
  ROSAT} HRI $0.1-2.4$ keV image (angular resolution of 5\arcsec). For
the HRI image, Arp (\cite{arp96}) evaluated the shift of the position
from the optical to X-ray to be of at most 7\arcsec\,, equivalent to one
MECS pixel. Three sources were identified in the MECS image,
\object{Mrk~205} included. Being in the field of view of the PDS also
(1.4 degrees while the MECS field of view is 28\arcmin), we investigated if one of the
other two sources could contribute to the
\object{Mrk~205} PDS spectra. \\
In addition, one had to make sure that \object{NGC~4319} located at
40\arcsec\, from \object{Mrk~205} (and thus not resolved by this
instrument) was not emitting in the X-rays (we used the following J2000
coordinates for \object{Mrk~205}:
12h\,~21\arcmin\,~44.12\arcsec\,~+75\degr\,~18\arcmin\,~38.2\arcsec\, and
for \object{NGC~4319}:
12h\,~21\arcmin\,~44.07\arcsec\,~+75\degr\,~19\arcmin\,~21.3\arcsec\,,
these are located by circles on Fig.~\ref{Fig2}). We discuss those
points below.
\begin{enumerate}
\item According to the profile displayed in Fig.~\ref{Fig2} (right
  panel), \object{NGC~4319} is located in a zone in which half of the
  MECS peak count rate is received. Furthermore the shape itself does not
  allow to tell if \object{NGC~4319} is really detected (the break in the
  profile can be a feature of the MECS PSF).  It was not detected by {\it
    ROSAT} HRI (Arp \cite{arp96}) in the soft X-ray domain, but it could
  be Compton thick with emission only detected above 5-10 keV.  However,
  there are no radio sources (apart from {\object Mrk~205}) above 2.5 mJy
  within a radius of 1.5\arcmin\, of {\object Mrk~205} in the NVSS
  catalog (Condon et al.  \cite{condon}). Thus it is very unlikely that
  \object{NGC~4319} is an AGN;
\item We have fitted the 2001 MECS spectrum with a power law,
  extrapolated the model in the PDS domain and added the PDS data without
  fitting them but including an intercalibration constant MECS/PDS frozen
  to the recommended value of 0.88 (see Fiore et al. \cite{fiore}).  This
  is of course not taking into account any reflection component.
  Fig.~\ref{Fig2b} shows the resulting spectra along with the residuals
  of both data sets with respect to the MECS best fit model.  It is very
  unlikely that emission from \object{NGC~4319} contributes to the excess
  at 20 keV and matches the extrapolation of the MECS flux at high
  energies.
\item The other sources in the MECS are
  \object{XMMU~J122206.4+752613}\footnote{the source
    \object{XMMU~J122206.4+752613} corresponds to \object{MS~1219.9+7542}
    and to {\it ROSAT} \object{1RXS~J122214.2+752624}} which is a narrow
  emission line galaxy and \object{NGC~4291}, a non-active elliptical
  galaxy. The first has a MECS count rate 10 times smaller than
  \object{Mrk~205} while for the latter it is 100 times smaller.
  According to Fiore et al. (\cite{fiore}), those sources are certainly in
  the residual PDS count rate (estimated by comparing to the known MECS
  count rate) obtained in blank fields (Guainazzi \& Matteuzzi
  \cite{guai});
\end{enumerate}

\noindent
From these different arguments, we conclude that neither
\object{NGC~4319} nor one of the two other sources detected in MECS
contribute significantly to the X-ray spectrum of \object{Mrk~205}.

\subsection{Possible absorption in the spiral arms of
  \object{NGC~4319}}
Of type S$_{pec}$, \object{NGC~4319} is an extremely disrupted 
spiral galaxy (Arp \cite{arp96}) with an ISM very different from the 
Milky Way's ISM. The ISM of \object{NGC~4319} has been thoroughly
studied by various authors which have found Ca II K 
and Mg I b absorption features at the redshift of \object{NGC~4319} 
(Stockton et al. \cite{stockton2}). \\
\noindent
Bahcall et al. (\cite{bahcall}) succeeded in the detection of
Mg II doublet in \object{NGC~4319} and deduced an approximative H I
column density of about $3\times 10^{18}$ cm$^{-2}$ from the Mg II
column density, precised by Womble et al. (\cite{womble}) to be 
N(H I)$<$ $5.2\times 10^{19}$ cm$^{-2}$. \\
\noindent
The gas in the inter-arm region of \object{NGC~4319} shows a
higher grade of ionisation than in the Milky Way and the unusually 
low amount of HI (or Mg II) is not understood for this kind of
late-type galaxy.\\
\noindent
The X-ray absorption in the $0.5-2.0$ keV energy range is however
dominated by CNO K-shell absorption, but no evidence for an enhanced
abundance of heavy elements is reported in the literature. 
To conclude, we do not find evidence for absorption in the
\object{Mrk~205} data due to the presence of \object{NGC~4319} in
the line of sight.

\subsection{Variability}
\begin{figure}
\includegraphics[width=9cm]{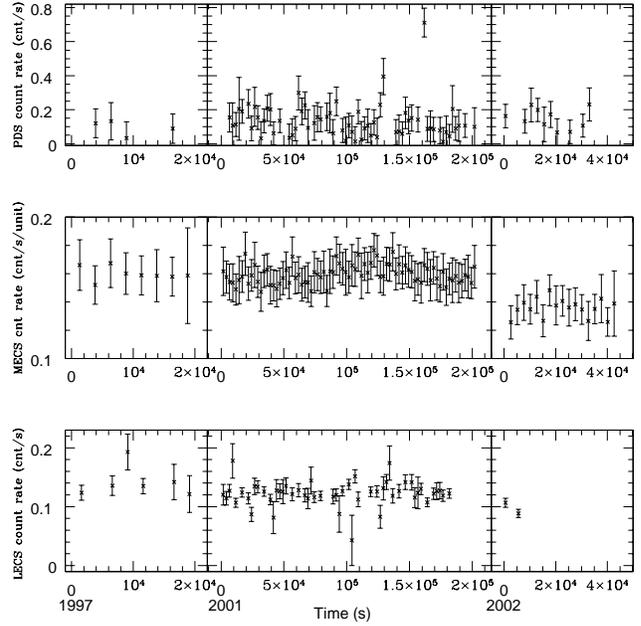} 
\caption{ Light curves of the 3 observations. Top panels: PDS $13-200$
  keV, middle panels: MECS $2-10$ keV (count rate per MECS unit),
  bottom panels: LECS $0.01-2$ keV}
\label{Fig3}
\end{figure} 
\begin{figure}
\includegraphics[width=9cm]{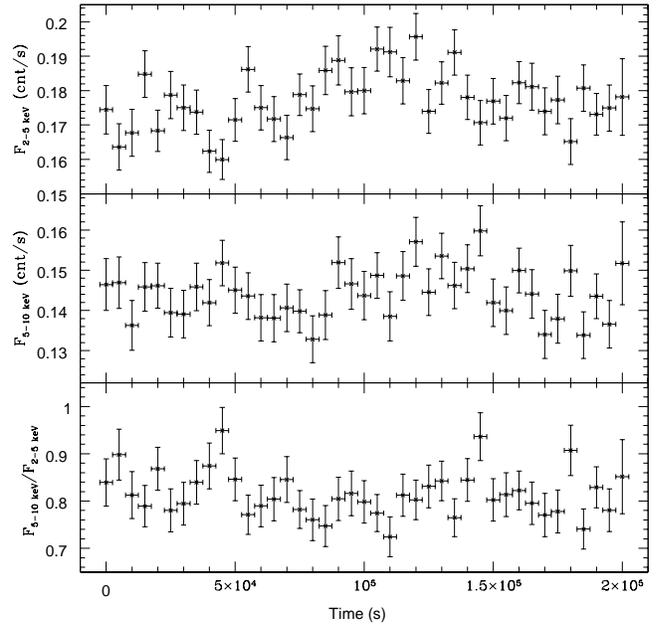}
\caption{ Observation of December 2001, MECS data. Top panel: $2-5$
  keV light curve, middle panel: $5-10$ keV light curve, bottom panel: 
hardness ratio 
}
\label{Fig4}
\end{figure}
\begin{figure}
\includegraphics[width=9cm]{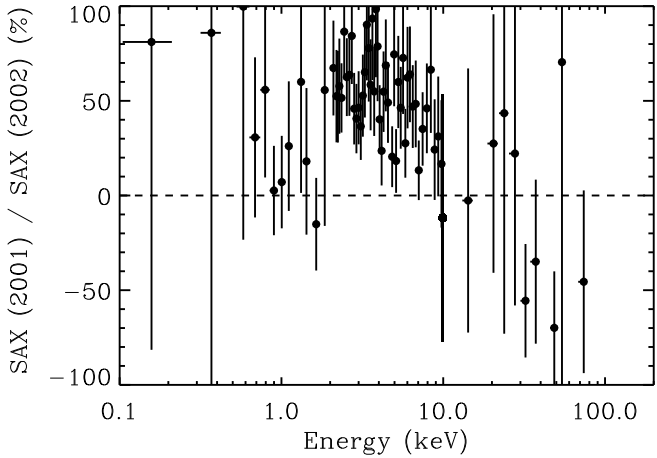}
\caption{ Spectral variations between 2001 and 2002 (ratio of the 2001 spectra to 
the 2002 spectra)}
\label{Fig5}
\end{figure}
\begin{figure}
\includegraphics[width=9cm]{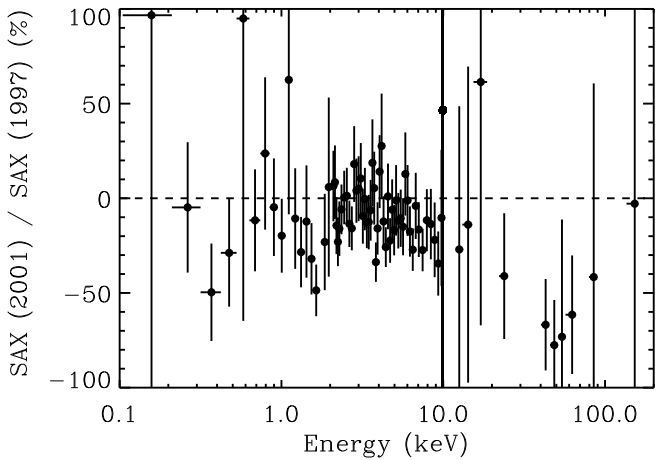}
\caption{ Spectral variations between 1997 and 2001 (ratio of the 2001 spectra to the
1997 spectra)}
\label{Fig6}
\end{figure}

We produced light-curves from the event files using Xronos 5.16 in a
single binning of 2500 s to ease intercomparison between the different
observations (Fig.~\ref{Fig3}). Note that 15 days separate the middle
panel of Fig.~\ref{Fig3} to the right panel. LECS and MECS light curves
of the statistically best observation (2001) were surprisingly different.
Cross-correlation analysis gave a weak correlation
coefficient between these two light curves.\\
From 1997 to 2001, the LECS and MECS count rates remained
constant. Since a MECS unit was lost between those 2 observations,
we have plotted in Fig.~\ref{Fig3} the MECS light curves for a 
single unit. The average PDS count rate seems to have increased by 
about 40\%, but this latter value is very inaccurate due to the few 
points in 1997. \\
From 2001 to 2002, the LECS average count rate seemed to have dropped
by about 16\% percent, but this latter value is based on a small
fraction of the observation time. The MECS average count rate dropped
by 15\%, while the PDS count rate remained constant. \\
From the December 2001 observation (Fig.~\ref{Fig3}, middle panels;
see also Fig.~\ref{Fig4}) one
sees that no significant variability was observed on very short
timescales (of the order of hours). On the other hand, long term
variability (on week and year timescale) can be quite
important (Fig.~\ref{Fig1}).\\

\noindent
In Fig.~\ref{Fig4}, we display the light curves $2-5$ keV and $5-10$ keV
along with the hardness ratio, for the longest observation (2001). The
$2-5$ keV energy band showed a weak minimum (10\% below the average flux)
around $4\times 10^{4}$ s which was not present in the $5-10$ keV range. 
The resulting hardness ratio thus showed an hardening around that
time. The differences in the two
light curves are certainly not due to $N_H$ variations as a value of
the order of $10^{20}$ cm$^{-2}$ is unlikely to change the spectra
above 2 keV.

\subsection{Model independent spectral analysis}
To study the spectral variations between the different observations,
we have divided
the 3 spectra (LECS, MECS, PDS) of the 2001 observation from the 3
spectra of the 2002 observation (Fig.~\ref{Fig5}). We have also divided
the 3 spectra of the 2001 observation from the 3
spectra of the 1997 observation (Fig.~\ref{Fig6}). We have taken 
the ratios for each energy
bins. One sees an important variation in the $2-10$ keV energy range between
the 2001 and 2002 observation which did not occur between the 1997 and
2001 observation, in agreement with Fig.~\ref{Fig3}. A rough comparison of Fig.~\ref{Fig5} and
Fig.~\ref{Fig6} shows that the LECS and PDS points may have
been enhanced as well in Fig.~\ref{Fig5}, but the signal-to-noise is not
good enough to conclude that the soft $0.1-2.0$ keV or hard $10-220$
keV bands varied significantly.

\begin{figure*}
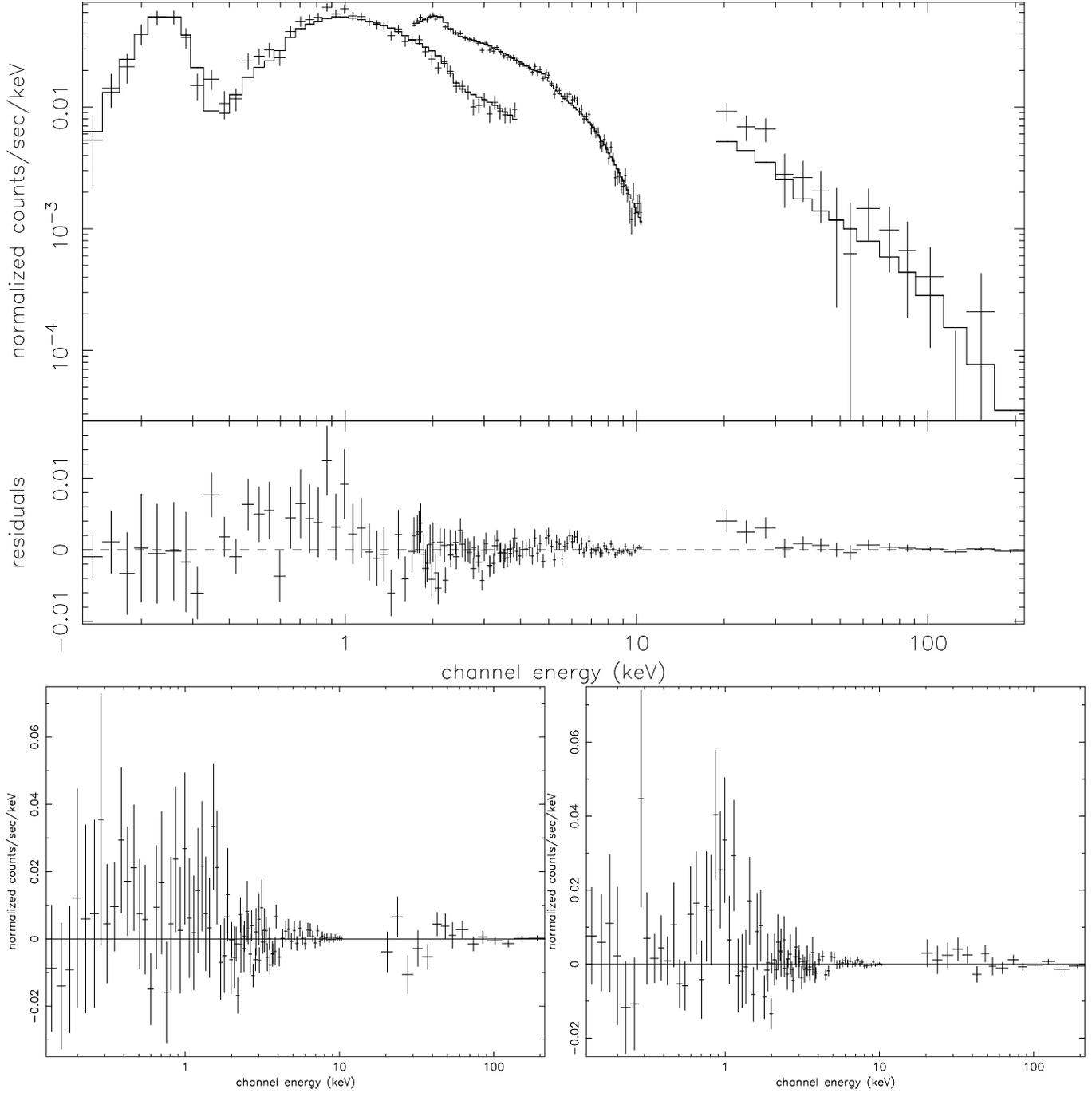

\includegraphics[height=17cm,angle=-90]{favre_fig8a.ps}
\includegraphics[height=9cm,angle=-90]{favre_fig8b.ps}\includegraphics[height=9cm,angle=-90]{favre_fig8c.ps}
\caption{ Upper panel: broad band (folded) spectrum of the 2001
  observation. The spectrum is fitted with a power law plus Galactic
  absorption model. Lower panel, left side: residuals after the fit of
  the 1997 observation with the same model. 
  Lower panel, right side: as above, for 2002}
\label{Fig7}
\end{figure*}

\subsection{Spectral analysis}
\begin{table*}[tb]
\caption{ Fits of the {\it BeppoSAX} data using a power law plus Galactic absorption model: best 
fit spectral parameters}\label{Tab2}
\begin{flushleft}
\addtolength{\tabcolsep}{-2pt}
\begin{tabular}{@{}lccc@{}}
\hline
\hline
\rule[-0.5em]{0pt}{1.6em}
Observation & $\Gamma$ & $N_H$ & $\chi^2$(dof), proba\\
\hline 
\rule[-0.5em]{0pt}{1.6em}
& & $10^{20}$ cm$^{-2}$&\\
\hline 
2001 & 1.87$^{+0.02} _{-0.03}$ & $<3.15$$^p$ &182.23(139), 0.008\\
1997 & 1.77$^{+0.07} _{-0.06}$ & $<4.08$$^p$ &106.88(99), 0.277\\
2002 & 1.72$^{+0.07} _{-0.07}$ & 3.69$^{+2.75} _{-0.79}$ &104.70(99), 0.328\\
\hline
\end{tabular}
\end{flushleft}
$^p$ pegged at Galactic value $N_H=2.90\times10^{20}$ cm$^{-2}$ 
\end{table*}
The spectral fitting was performed with XSPEC 11.2.0. In the following,
all errors are quoted at the 90\% confidence level for one interesting
parameter. We have used the recommended energy bounds (LECS:~$0.12-4.0$
keV, MECS:~$1.65-10.5$ keV, PDS:~$15-220$ keV), and the intercalibration
``constants'' were free to vary in the intervals $0.7-1.0$ and
$0.82-0.90$ for LECS/MECS, PDS/MECS respectively
(Fiore et al. \cite{fiore}).\\
\\
The upper panel of Fig.~\ref{Fig7} represents a fit of the 2001 data with
a power law model modified by interstellar photoelectric absorption. 
The photoelectric absorption cross
sections used were those of Morrison and McCammon (\cite{morrison}). \\
The lower panels of Fig.~\ref{Fig7} show the fits residuals with the same
model for the 1997 and 2002 observations respectively. The best fit
spectral parameters found are presented in Table~\ref{Tab2}.
\noindent
The fit of the 2001 observation is poor with an excess in the $15-40$ keV
range as well as in the soft X-rays, up to about 1 keV, finally a small
feature around 6 keV is present.\\
For the 1997 and 2002 observations, the fits are already acceptable 
with this simple
model for the continuum. In the 2002 observation, a line-like feature is
clearly detected around 1 keV while a small excess in the range $15-40$
keV may be present. The signal-to-noise ratio of the continuum
is not good enough to accurately constrain more complex physical
models.

\subsection{Spectral analysis: reflection component}
\begin{figure}
\includegraphics[height=9cm,angle=-90]{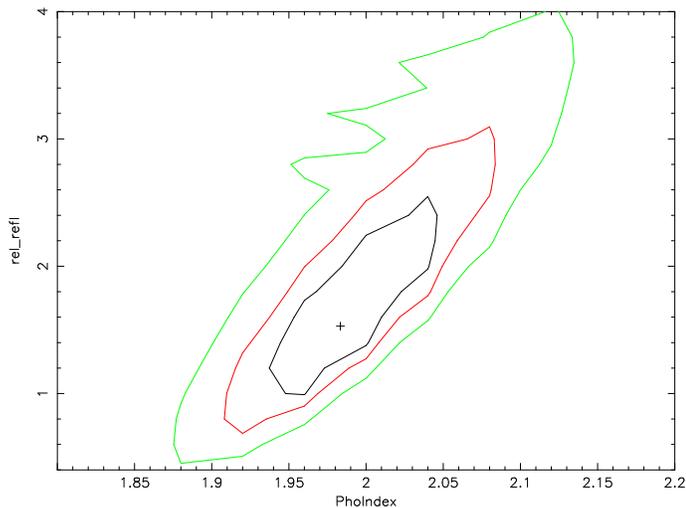}
\caption{ Contour plot $R$-$\Gamma$ after the fit of the 2001
  observation with the Compton reflection from cold material model plus an Fe K$\alpha$ line}
\label{Fig10}
\end{figure}

\begin{figure}
\includegraphics[height=9cm,angle=-90]{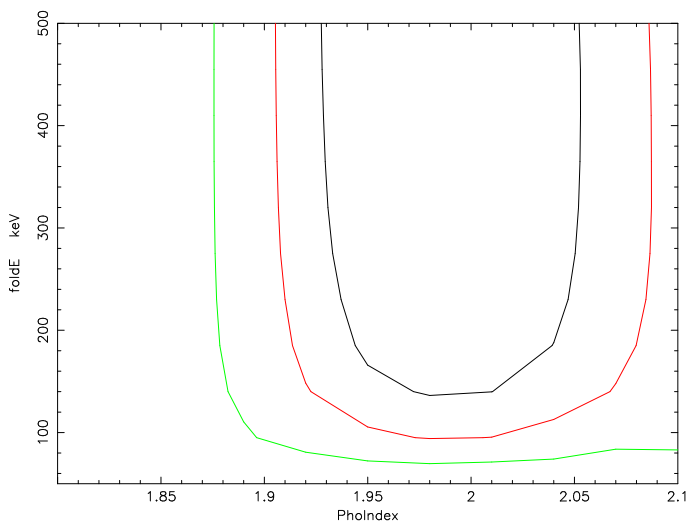}
\caption{ Contour plot $E_c$-$\Gamma$ after the fit of the 2001
  observation with the Compton reflection from cold material model 
plus an Fe K$\alpha$ line}
\label{Fig11}
\end{figure}

We started a more detailed analysis by fitting the 2001 observation
data above $\sim$ 2 keV, i.e. using MECS and PDS data only. As $N_H$
was not constrained in this case, we fixed it to the Galactic value. 
We tried 2 different models for the continuum shape. \\
\noindent
The first was a cut-off power law plus a Compton reflection from dense,
neutral matter ({\sc pexrav}: Magdziarz \&
Zdziarski \cite{magdz}). This model is meant to describe 
the effects of Compton
scattering of photons from an X-ray source associated with the inner
parts of the accretion disk. The spectrum of reflected X-ray was computed
as a function of disk inclination using the transfer function of
Magdziarz \& Zdziarski (\cite{magdz}). The free parameters of this model
are the spectral index $\Gamma$ of the primary power law, the folding
(upper cutoff) energy $E_c$ of the primary X-ray spectrum and the
normalization of the reflection component $R$. In our fits, the inclination
angle of the disk was frozen to 30 degrees as it is usually admitted for
Seyfert~1 galaxies, while the abundances of Fe and other heavy elements
were fixed to the solar values.
\\
The second model was a more physical model of thermal Comptonisation for
a disk-corona configuration in a slab geometry ({\sc compha}: Haardt 
\cite{haar}; Haardt et al. \cite{haardt}). The table model for
XSPEC\footnote{http://pitto.mib.infn.it/$\sim$haardt/ATABLES} includes both
the primary continuum and the reflection.  The reflection component was
computed following White et al. (\cite{white}) and Lightman \& White
(\cite{lightman}) assuming reflection on neutral matter and a constant
spectral shape (angle-averaged) for the reflected photons.  The free
parameters of the model are the corona temperature normalized to the
electron mass $kT_e/m_ec^2$,
the corona optical depth $\tau$ and the normalization of the reflection
component $R$ (without the comptonisation effects). In our fits, the
temperature $kT_{bb}$ of the disk (which is assumed to radiate like a
single black-body) was frozen to 10 eV while the inclination angle of the
disk was frozen to 30 degrees.
\\
Both continuum models were completed by the addition of a Gaussian
emission line to model the Fe~K$\alpha$ line with a width of 0.1 keV
frozen (a more detailed analysis on the Fe line parameters is made
below).  The best fit spectral parameters are given in Table~\ref{Tab3}.
\begin{table*}[tb]
\caption{ Reflection component: fits of the MECS and PDS data ($1.65-220$ keV). A stands for
  photoelectric absorption, PL for power law, {\sc pexrav} for the Compton
  reflection from cold material and {\sc compha} for the thermal 
comptonisation 
  model. $\Gamma$ is the photon index of  the primary power law, $E_c$ is
  the folding (upper cutoff) energy, $R$ is  
  the normalization of the reflection component, $kT_e/m_ec^2$ is the corona
  temperature (normalized to the electron mass), $\tau$ is the corona 
optical depth,
  $E_{\rm Fe}$ is the Fe~K$\alpha$ 
  line energy (rest frame) and $EW_{\rm Fe}$ its equivalent width.
    As $N_H$ is not constrained above 1.65 keV, we fixed it to the
  Galactic value. In addition, the results of a fit of the {\it
  XMM-Newton} data above 2 keV are presented}\label{Tab3}
\begin{flushleft}
\addtolength{\tabcolsep}{-2pt}
\begin{tabular}{@{}llcccccccc@{}}
\hline
\hline
\rule[-0.5em]{0pt}{1.6em}
Obs. & Model  & $\Gamma$ & $E_c$ & $R$ & $kT_e/m_ec^2$ & $\tau$ &$E_{\rm Fe}$& $EW_{\rm Fe}$ & $\chi^2$(dof), prob.  \\ 
\hline
\rule[-0.5em]{0pt}{1.6em}
&&& keV &&&& keV & eV &\\
\hline
{\it SAX} 2001&A+PL  & 1.83$^{+0.03} _{-0.03}$ &-&-&-&-&-&-&137.47(93), 0.002\\ 
&A+PEXRAV+G  & 2.00$^{+0.03} _{-0.03}$ & 487.17$^{+u} _{-304.40}$ & 1.74$^{+0.25} _{-0.32}$ &-&-& 6.51$^{+0.42} _{-0.55}$ & 71.30& 100.38(89),
0.192 \\
&A+COMPHA+G &-&-&3.22$^{+0.99} _{-1.35}$&0.32$^{+0.04} _{-u}$&0.15$^{+0.33} _{-0.08}$&6.52$^{+0.18} _{-0.18}$&70.30&100.39(89), 0.192\\  
\hline 
{\it SAX} 1997&A+PL  & 1.74$^{+0.09} _{-0.08}$ &-&-&-&-&-&-& 48.36(53), 0.655\\ 
&A+PEXRAV+G  & 1.77$^{+0.70} _{-0.13}$ & 240.00$^f$ & $<8.78$ &-&-&6.40$^f$& $<$1&48.16(51), 0.587\\
&A+COMPHA+G  &-&-&2.50$^f$&0.48$^{+0.04} _{-u}$&0.05$^{+u} _{-u}$&6.40$^f$&$<$1&47.66(51), 0.607\\ 
\hline
{\it SAX} 2002&A+PL  & 1.69$^{+0.07} _{-0.07}$&-&-&-&-&-&-& 49.91(53), 0.595\\ 
&A+PEXRAV+G  & 1.80$^{+0.25} _{-0.16}$ & 228.30$^{+u} _{-193.00}$ & $<3.47$ &-&-&6.40$^f$&34.90& 46.89(50), 0.599 \\
&A+COMPHA+G &-&-&2.43$^{+4.41} _{-1.97}$&0.37$^{+0.13} _{-0.09}$&0.19$^{+u} _{-u}$&6.40$^f$&38.90&46.27(50), 0.624\\ 
\hline
{\it XMM} 2000&A+PEXRAV+G  & 1.73$^{+0.14} _{-0.14}$ & 500.00$^{f}$ & 0.48$^{+1.09} _{-0.48}$ &-&-&6.40$^f$&75& 75.24(54), 0.029 \\
\hline
\end{tabular}
\end{flushleft}
$^f$ parameters fixed in the fit \\
$^u$ unbounded parameters
\end{table*}
\noindent
The fit of the observation made in 2001 is improved by using 
{\sc pexrav} without cutoff
($\Delta\chi^2$=29.51 for 1 more parameters; F-test value 25.20 and
probability $2.57\times 10^{-6}$). It is slightly improved
when including the cutoff ($\Delta\chi^2$=1.97 for 1 more parameters;
F-test value 1.70 and probability 0.196)   
and the line ($\Delta\chi^2$=5.60 for 2 more parameters; F-test value 
2.48 and probability $8.93\times 10^{-2}$).
The solid angle subtended by the reflector is poorly
constrained (Fig.~\ref{Fig10}). Figure~\ref{Fig11} shows that the cutoff
upper limit is not constrained.\\
\noindent
The $\chi^2$ is not significantly improved by {\sc pexrav} for the two other 
observations (Table~\ref{Tab3}). We thus cannot constrain
the presence of this component in 1997 and 2002.
\\
\\
\noindent
The fit of the December 2001 observation is improved by using 
{\sc compha} when compared to a power law
($\Delta\chi^2$=30.11 for 2 more parameters; F-test value 12.80 and
probability $1.30\times 10^{-5}$). It is slightly improved 
by including the Fe line ($\Delta\chi^2$=7.17 for 2 more parameters; 
F-test value 3.19 and probability $4.61\times 10^{-2}$).
\\
\\
\noindent
To summarize, fitting with the thermal comptonisation model {\sc
  compha} does not improve the fit of
the 2001 observation in comparison to the Compton
  reflection from cold material
{\sc pexrav} fits. 

\subsection{Spectral analysis: soft excess}
We now consider the broad band 2001 spectrum, adding the LECS 0.12-4.0
keV data to the MECS and PDS data.  Freezing all the parameters of the
best fit {\sc pexrav} model above 2 keV and fitting again the broad band
spectrum, we get an $N_H=(3.46^{+0.26} _{-0.22})\times 10^{20}$
cm$^{-2}$, and a $\chi^2$ of 138.49 for 142 degrees of freedom
(dof). We found no clear evidence
for a soft excess of photons in the LECS, even if the intercalibration
constant is set to its lowest value of 0.7.  To have an upper limit on a
soft excess component, we add a black body to our model, fixing its
temperature to the one obtained by Reeves et al. (\cite{reeves2}) with the
{\it XMM-Newton} data, i.e. $kT=120$ eV. 
We found an $N_H=(3.60^{+0.36}_{-0.29})\times 10^{20}$ cm$^{-2}$ and 
a $\Delta\chi^2=1.73$ for 1 more parameter;
F-test value 1.78 and probability $0.184$. 
The soft excess is here much weaker ($\sim$ 3\% of the emission over the
range 0.5-2.0 keV) than it was claimed in the {\it XMM-Newton}
observation ($\sim$ 15\% of the emission in the range 0.5-2.0 keV). 
 
\begin{figure*}
\includegraphics[height=9cm,angle=-90]{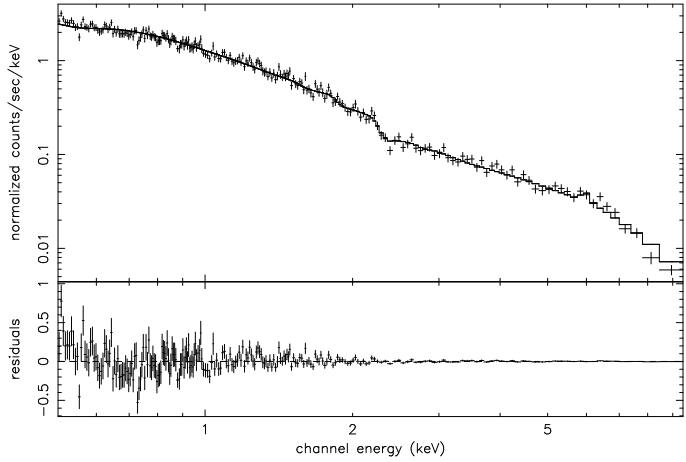}
\caption{ {\it XMM-Newton} 2000 data, fit with the Compton
  reflection from cold material model}
\label{fig11b}
\end{figure*}

\begin{table*}[tb]
\caption{ Soft excess: fits of the LECS, MECS, PDS data and comparison 
with {\it XMM-Newton} data. A stands for photoelectric absorption, 
{\sc pexrav} for the Compton reflection from cold material, G for a 
Gaussian line and BB for a black body. We first consider the spectra
above 2 keV, freezing all the parameters of the continuum ($\Gamma$,
$R$, $E_c$) to their best fit value of Table~\ref{Tab3}. We then
add the data below 2 keV, add a soft excess component, and fit again 
the broad band spectrum}\label{Tab4}
\begin{flushleft}
\addtolength{\tabcolsep}{-2pt}
\begin{tabular}{@{}llcccc@{}}
\hline
\hline
\rule[-0.5em]{0pt}{1.6em}
Obs.&Model & $N_H$ & $kT_{\rm{BB}}$ & $\chi^2$(dof) & proba  \\    
\hline
\rule[-0.5em]{0pt}{1.6em}
 && 10$^{20}$ cm$^{-2}$ & eV && \\
\hline
{\it BeppoSAX} 2001&A+PEXRAV+G+BB & 3.60$^{+0.50} _{-0.31}$ & 100.00$^{+u} _{-99.90}$ & 137.40(140) & 0.546\\
\hline
{\it XMM} 2000&A+PEXRAV+G+BB    & $<3.14^p$ & 164.86$^{+4.85} _{-8.57}$ & 325.09(231)
&$<0.001$ \\
\hline
\end{tabular}
\end{flushleft}
$^u$ unbounded parameters\\
$^p$ pegged at Galactic value $N_H=2.90\times10^{20}$ cm$^{-2}$ 
\end{table*}
\noindent
We re-analysed the {\it XMM-Newton} PN data to see whether the soft
excess component was as important as claimed by Reeves et al.
(\cite{reeves2}) when the reflection component was present. Indeed,
Reeves et al. (\cite{reeves2}) discussed the soft excess component
modeling the continuum with a power law while Page et al. (\cite{page})
used the Compton reflection model but stopped their analysis at 3 keV. As
before, we deduced the parameters of the continuum fitting the data above
2 keV (Table~\ref{Tab3}). The {\sc pexrav} model produced a 
$\chi^2$ of 75.24 (54 dof) with a realistic reflection parameter $R$
of 0.48. We then fixed the continuum and included the 0.5 to 2.0 keV
points. The residuals showed a clear excess below 2 keV. Adding a
black body with a temperature of 120 eV did not produce a good fit
($\chi^2=463.29$ for 232 dof) but, when this parameter was released,
the fit was much better and the temperature significantly increased
(see Table~\ref{Tab4} and Fig.~\ref{fig11b}). We did not find an 
unusual value of $N_H$ (Fig.~\ref{Fig1}) anymore.\\
The contribution of the soft excess component is thus more important
than previously claimed: it represents $\sim$ 23\% of the emission in
the range 0.5 to 2.0 keV.

\noindent
However, the black body parameters do show suspicious values
meaning that the emission is probably not a simple black body.

\subsection{Spectral analysis: Fe K$\alpha$ line}
To assess the importance of an Fe K$\alpha$ line in our data, we
concentrate on
the MECS data. In this case, $N_H$ is not constrained, we thus fixed it
to the Galactic value.  Using only the power law model, it is
clear from the residuals in the lower panel of Fig.~\ref{Fig7} that a line
feature is detected in the 2001 observation, while it does not seem to be
present in the 2002 observation (same figure). The
statistics of the 1997 observation is too poor to give conclusive results.\\
On the other hand, after fitting the continuum of the 2001 data with
the Compton 
reflection model, the residuals do not show a clear evidence for a 
line feature anymore.
\\
\\
We thus concentrate on the 2001 best fit observation and add a 
Gaussian emission
line. The fit is improved ($\Delta\chi^2=6.07$ for 1
more parameter; F-test 5.39 and probability $2.29\times 10^{-2}$)
with a line at 6.53 keV consistent with neutral 
Fe~K$\alpha$ and weakly ionized
species. Note that the line is not resolved as $\sigma_{Fe}<500$ eV is 
smaller than the MECS energy resolution.\\
Both Compton reflection from cold material and thermal
comptonisation model give a line energy of 6.53 keV
but the latter gives a slightly larger equivalent width. 
According to George \& Fabian (\cite{geofab}), if a reflection is
present, it has to be associated with an Fe line. In the neutral case,
they computed an equivalent width of the order of 1.7 keV with respect
to the reflection or $\sim$ 150 eV with respect to the continuum, for
$R=1$ (see also Zdziarski et al. \cite{zdziarski}). Our measures
of the line equivalent width and $R$ are inconsistent with the
predictions of George \& Fabian (\cite{geofab}). We shall discuss in
Section~4 a possible reason to this deviation.
\\
\\
\noindent
Unlike Reeves et al. (\cite{reeves2}) and Page et al.
(\cite{page}), no residuals are found at $\sim$ 7 keV, but the line
equivalent width and centroid suggest that it could also be a blend of a
neutral narrow component and a weakly ionized one. Adding a second
Gaussian to the model while freezing the line energies to 6.4 keV and 6.7
keV rest frame following Reeves et al. (\cite{reeves2})
was not significant and neither was the relativistically-broadened
accretion disk line
model (Fabian et al. \cite{fabian}) which was preferred by Reeves et
al. (\cite{reeves2}).
\begin{table*}[tb]
\caption{ Fe K$\alpha$ line: fits of the 2001 observation, MECS data
  only. A stands for photoelectric absorption (fixed to the Galactic
  value), {\sc pexrav} for the Compton reflection from cold material, 
{\sc compha} for the thermal comptonisation model and G for Gaussian
  (line energies are rest frame). $R$, $E_c$, $kT_e/m_ec^2$ and $\tau$ were 
fixed to their best fit values of Table~\ref{Tab3}}\label{Tab5}
\begin{flushleft}
\addtolength{\tabcolsep}{-2pt}
\begin{tabular}{@{}lcccccc@{}}
\hline
\hline
\rule[-0.5em]{0pt}{1.6em}
Model & $\Gamma$ & $E_{\rm Fe}$ & $\sigma_{\rm Fe}$ & $EW_{\rm Fe}$ &$\chi^2$(dof) & prob\\
\hline
\rule[-0.5em]{0pt}{1.6em}
&& keV&eV & eV && \\
\hline
A+PEXRAV&1.99$^{+0.03} _{-0.02}$&-&-&-&92.81(78)& 0.121\\
A+PEXRAV+G&1.99$^{f}$&6.53$^{+0.23} _{-0.19}$&$<447.60$& 56.80&86.74(77)&0.210\\
A+COMPHA&-&-&-&-&95.53(79)&0.099\\
A+COMPHA+G&-&6.53$^{+0.16} _{-0.16}$&$<314.82$&71.90&86.25(77)&0.220\\
\hline
\end{tabular}
\end{flushleft}
$^f$ parameters fixed in the fit 
\end{table*}
\noindent
For the 2002 observation, the residuals were not improved when using
the Compton reflection model and thus were identical to Fig.~\ref{Fig7}.

\subsection{Line feature at 1 keV}
\begin{figure}
\includegraphics[height=9cm,angle=-90]{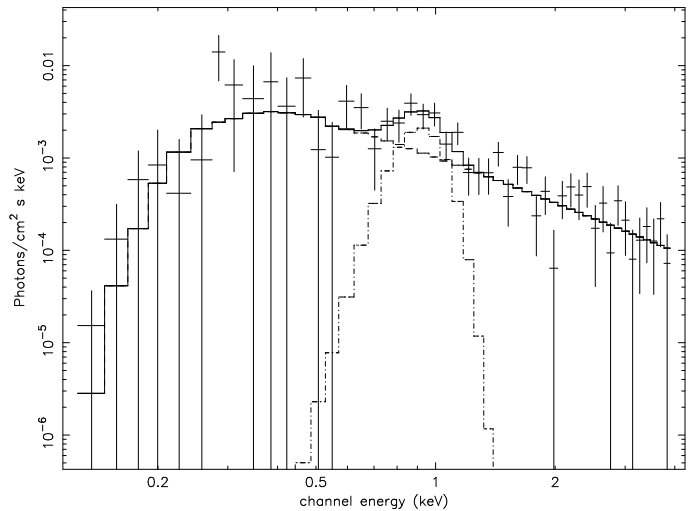}
\caption{ Observation taken in 2002, LECS data only, unfolded spectrum
  fitted with a power law plus Galactic absorption and a Gaussian
  component centered at an energy of about 1 keV
}
\label{Fig18}
\end{figure}

The spectrum of the 2002 observation contains a narrow emission line
feature centered at about 1 keV (see Fig.~\ref{Fig7}, lower right panel).
We have fitted a model composed of a power law plus Galactic absorption
and a Gaussian component (Fig.~\ref{Fig18}) and found the following
results. The line is centered at $0.98^{+0.08} _{-0.34}$ keV (quasar rest
frame) has a width of $0.11^{+0.33} _{-0.11}$ keV (not resolved; the
LECS energy resolution at 1 keV is $\sim$ 0.2 keV) and
an equivalent width
of 518 eV. The addition of the line significantly improves the fit at better
than the 98\% confidence level according to the F-test. Note
however that even if this simple model allows us to estimate the line 
centroid energy and total flux, it is physically unrealistic as it is
likely that this line results from a blend of several emission lines,
predominantly from ionized species of Ne and possibly from Fe L shell.
\\
A similar feature was also found in the {\it ASCA} spectrum of the quasar
\object{PG~1244+026} (Fiore et al. \cite{fiore5}; George et al.
\cite{george}) as well as in the {\it ASCA} spectrum of the Seyfert~1
galaxies \object{TON~S180} (Turner et al. \cite{turner1}) and
\object{Akn~564} (Turner et al. \cite{turner2}). \\
We searched the {\it XMM-Newton} data for this feature but did not find a
clear evidence for it (the equivalent width reached a maximum of 10 eV).
\section{Discussion-Conclusion}
\noindent
{\bf The nature of \object{Mrk~205}.} 
It has been noted that, in contrast to the Seyfert~1 case, the 
detection of Fe K$\alpha$ line in quasars was more seldom and 
that if detected, the line energy
was around 6.7 keV instead of 6.4 keV (Reeves \& Turner \cite{reeves1}).
It was thus suggested that sensitive instruments like the EPIC camera on
{\it XMM-Newton} could observe the Fe K$\alpha$ emission of quasar in
detail. From our X-ray observations of \object{Mrk~205}, and from the
luminosity we obtained from them, we suggest to be careful in classifying
this object. Its luminosity is found in the region in which it is difficult
to choose for a Seyfert~1 or for a quasar. It seems thus difficult to 
generalize any findings
made on this object to a whole class.
\\
\\\noindent
{\bf The soft excess variability.} 
The soft excess seems to be variable in this source 
although a very strong soft excess was never reported.
It seems to be absent/not detected in 1997, strong in 2000, weak 
in 2001 and absent
again in 2002 (thus varied in a 2 weeks timescale). The variability of
the soft excess cannot be associated with variability of a cold absorber
as we have shown in Fig.~\ref{Fig1} and Section 3 that the absorbing
column density deduced from spectral
modeling was roughly constant over all the observations. \\
Note that the soft excess component measured here should not be
associated with direct thermal emission from the accretion disk because
its temperature is much too high.
\\
\\\noindent
{\bf Thermal comptonisation.} The historic variations of the photon index
can be explained in the framework of thermal comptonisation. 
\object{Mrk~205} showed large $2-10$ keV flux variations
between the 1997 (high flux state), 2000 (low flux state),
2001 (high flux state) and 2002 (low flux state) observations which do
not seem to be correlated with variability of the soft excess. \\
However, in thermal comptonisation processes, the index of the comptonised
component is inversely correlated with the ratio of the photon seed flux
to the comptonised power law flux. We can test this hypothesis if we
assume that the soft excess strength is proportional to the strength of
the soft photon field (Walter \& Fink \cite{walterfink}; Walter \& 
Courvoisier \cite{walter1}).
The index for 2000 was $1.73\pm 0.02$ for a ratio of 23\% while for 2001
the index was $2.00\pm 0.03$ for a ratio of 3\%. This is thus in
disagreement with our expectations.
\\
On the other hand, using the {\sc compha} model to estimate the
UV flux, we found that an increase of the photon index goes with a
larger UV flux, in agreement with simple thermal comptonization
model expectation. Similar conclusions were found by Petrucci et
al. (\cite{pet00}) for the Seyfert~1 galaxy \object{NGC~5548}. 
The contradiction between the soft X-ray and UV variations is not very 
surprising. Indeed the soft excess component is certainly linked to
the UV bump through complex radiative transfer (e.g. comptonization
in the warm (few keV) upper layer of the accretion
disk). Consequently, its expected spectral behavior is not
straightforward and may be at odds from the zeroth order expectations.
\\
\\\noindent
{\bf Constraints on the reflecting material.}
The nature of the reprocessing 
medium can be the
accretion disk or some obscuring, distant material. If the Fe K$\alpha$
line is produced in the inner parts of the accretion disk, its profile 
should be
considerably broader, with very distinct asymmetries due to Doppler and
gravitational redshift. This is not possible to test with the {\it
  BeppoSAX} data while Page et al. (\cite{page}) report that they did not
find evidence for it in the {\it XMM-Newton} data either.\\
The strength of the Compton reflection $R$ is here
systematically higher than what was found in
Seyfert galaxies ($R\simeq 0.8\pm 0.2$, Nandra \& Pounds 
\cite{nandra}), while on the contrary, the iron line equivalent width
EW is smaller than the mean EW value. It is worth noting however that
our measures of $R$ are consistent with $R<1$ at 90\% confidence, 
see Fig.~\ref{Fig10}.\\
The inconsistency between the best fit values of $R$
and the line EW may then occur by chance and the reflecting material
in \object{Mrk~205} may simply cover a solid angle $<2\pi$ as seen
from the X-ray source. The line energy then also suggests the material
to be neutral. A low Fe abundance will make the line weaker and the 
Compton reflection hump stronger (as observed) by reducing the opacity
above the Fe K edge (George \& Fabian \cite{geofab}).\\
Given these constraints a simple solution could be that the
reprocessing is done in a distant material e.g. the dust torus.\\
\\
However, we cannot rule out the presence of close mildly ionized
material. Indeed Nayakshin \& Kallman (\cite{naykal}) have shown 
that in the case of relatively strong illumination by local X-ray 
flares, the upper layer of the reflecting material may become strongly 
ionized. In this case the reflection features are mainly produced by
the neutral depth layer and the expected line energy is 6.4 keV as we 
observed. The line EW is however smaller than the one expected in the 
case of neutral reflector since the line may be partly comptonized in 
the hot skin. But clearly this skin cannot be highly ionized since in 
this case it would act like a perfect mirror and the reflection hump
would be suppressed contrary to what we observe. Lines in the soft
band are also expected in the case of a mildly ionized medium and it 
may explain the poor fit of the soft excess with a simple black body 
component.
\noindent
\\  
\\  
{\bf The 1 keV line.} The spectrum of January 2002 contains a narrow
emission feature centered at an energy of $\sim$ 1 keV which was 
also found in at least 3 other AGN. This feature could be explained 
by a blend of several emission lines, predominantly from ionized species of 
Ne and possibly from Fe L shell. The line emission could be a result
of reflection in photoionized matter. It could also be resulting 
from recombination in
an optically thin thermal plasma.
\begin{acknowledgements}
PF acknowledges a grant from the Swiss National Science Foundation. This research has 
made use of the NASA/IPAC Extragalactic Database (NED) which is operated by the Jet 
Propulsion Laboratory, California Institute of Technology, under contract with the 
NASA. This research has also made use of the 
SIMBAD database, operated at CDS, Strasbourg, France.
\end{acknowledgements}

\end{document}